\documentclass[aps,twocolumn,showpacs,preprintnumbers, floats]{revtex4}
\usepackage{amsfonts}
\usepackage{amsmath}
\usepackage{graphicx}
\usepackage{dcolumn}
\usepackage{bm}

\begin{document}

\preprint{}
\title{Polar domains in lead titanate films under tensile strain}
\author{G. Catalan$^{1,}$$^{3}$}
\email{gcat05@esc.cam.ac.uk}
\author{A. Janssens$^{2}$, G. Rispens$^{1}$, S. Csiszar$^{1}$, O. Seeck$^{4}$%
, G. Rijnders$^{2}$, D.H.A. Blank$^{2}$}
\author{B. Noheda$^{1}$}
\email{b.noheda@rug.nl, Corresponding author}
\affiliation{$^{1}$Materials Science Centre, University of
Groningen, Groningen 9747AG, The Netherlands}
\affiliation{$^{2}$MESA+ Institute for nanotechnology, Twente
University, Enschede 7500 AE, The Netherlands}
\affiliation{$^{3}$Department of Earth Sciences, University of
Cambridge, Cambridge CB2 3EQ, United Kingdom}
\affiliation{$^{4}$HASYLAB- DESY, Notkestr. 85, D-22603 Hamburg,
Germany}
\date{December 20, 2005}

\begin{abstract}
Thin films of PbTiO$_{3}$, a classical ferroelectric, have been
grown under tensile strain on single-crystal substrates of
DyScO$_{3}$. The films, of only 5nm thickness, grow fully coherent
with the substrate and show no crystallographic twin domains, as
evidenced by synchrotron x-ray diffraction. A mapping of the
reciprocal space reveals intensity modulations (satellites) due to
regularly-spaced polar domains in which the polarization appears
rotated away from the substrate normal, characterizing a low
symmetry phase not observed in the bulk material.
This could have important practical implications since these
phases are known to be responsible for ultrahigh piezoelectric
responses in complex systems.

\end{abstract}

\maketitle

Ferroelectrics are dielectric materials with a permanent switchable
polarization. Ferroelectricity as a critical phenomenon is of broad
fundamental interest. Theoretical and experimental research
of ferroelectric thin films is recently attracting a lot of attention not
only because of their promising technological applications in novel geometries but also because of the new
fundamental understanding of ferroelectricity at the
nanoscale\cite{Dawber}.

Perovskite ferroelectric thin films are strongly affected by
epitaxial strain. Clamping between the film and the substrate onto
which it is deposited is known to induce shifts in critical
temperatures, increase tetragonality and polarisation, and change
the order of the phase transition \cite{Rossetti91, Choi04, Haeni04,
He05,Basceri97, Pertsev98}. Some theoretical works have also
predicted other interesting possibilities, such as changing the
polar symmetry \cite{Pertsev98, Dieguez05, Bungaro04}, engineering
the domain configuration \cite{Speck94, Li01} and, for unscreened surface charges, inducing polar
rotations across the thickness of the films \cite{Kornev04}.

The most interesting predictions of the stability diagrams for
epitaxial tetragonal perovskite ferroelectrics tend to be those for tensile
in-plane strain \cite{Pertsev98, Bungaro04,Dieguez05, Speck94,
Li01}, as compressive strains simply induce an enhancement of
tetragonal out-of-plane polarization. However, experimental
verification of such predictions is difficult. This is partly
because most available perovskite single-crystal substrates
(SrTiO$_{3}$, LaAlO$_{3}$, NdGaO$_{3}$) have lattice parameters that
are either very close to, or smaller than, those of typical
perovskite ferroelectrics such as PbTiO$_{3}$ or BaTiO$_{3}$. At the
other end of the spectrum, substrates with bigger lattice
parameters, such as perovskite KTaO$_{3}$, have so large a mismatch
that the strain is rapidly relaxed through dislocations or twin
formation\cite{Kwak,Stemmer}. Fortunately, the recent development of
new perovskite single crystals from the family of the scandates
\cite{Choi04, Haeni04, Biegalski05} bridges the gap, thereby
widening the scope for strain tailoring of the films and allowing us
to explore experimentally some of the above-mentioned theoretical
predictions.

We have grown fully coherent thin films of the archetypal perovskite
ferroelectric PbTiO$_{3}$ (PTO) under biaxial tensile strain and
analyzed the reciprocal space using synchrotron x-ray diffraction.
The results show that polarization rotation is induced in these
films. This new polar state of PTO is consistent with the so-called
\textit{ac} -polarization tilted in the (010) plane- and \textit{r}
 -polarization tilted in the (110) plane- phases. Neither symmetry
exists in the parent bulk compound \cite{Kobayashi83}. These
findings could have important technological implications since
similar low symmetry phases, very rare in ferroelectrics, are believed to be responsible for the
unusually large responses observed in the ultrahigh piezoelectrics,
such as PbZr$_{1-x}$Ti$_{x}$O$_{3}$ and related systems\cite{Noheda00,
Bellaiche00}. The results thus confirm experimentally that new polar
states that do not exist in bulk can be made accessible in epitaxial
films through careful tuning of strain, thickness and electrode
\cite{Pertsev98,Dieguez05}.

Experimental studies of ferroelectricity often require either the
use of electrodes, which themselves introduce important changes in
the thin films \cite{Junquera03, Kornev04}, or optical methods,
unfeasible when the optical thickness of the films is smaller than
the wavelength. However, in a series of seminal papers, Streiffer
and co-workers have shown that polar stripe domains can be detected
as satellite peaks in x-ray diffraction patterns, and that this
technique can be used to measure the domains size and polar
orientation \cite{Streiffer02, Fong04} even in films as thin as 3
perovskite unit cells \cite{Fong04}. This technique is therefore
ideally suited for the study of polar symmetry in very thin films
where very few others (see ref. \cite{Triscone05}) can be
implemented.

In this study we have deposited 5nm thin films of PTO, using
RHEED-assisted Pulsed Laser Deposition, at 570$^{o}$C and with an
oxygen background pressure of 0.13 mbar (further details of the
growth are published elsewhere\cite{Janssens06}). The films were
grown either directly on [110]-DSO crystals, or on [110]-DSO with a
5nm buffer layer of the conductive perovskite SrRuO$_{3}$ (SRO),
deposited at 600$^{o}$C\cite{Rijnders}. Bulk PTO has a tetragonal
structure at room temperature, with lattice parameters a=b=3.89{\AA
} and c=4.14{\AA } \cite{Shirane50}. The lattice structure of
DyScO$_{3}$ (DSO) is orthorhombic and [110]-oriented substrates are
expected to have nearly-squared in-plane lattice with a=
3.944-3.948{\AA } and b=3.943-3.945{\AA } at right
angles\cite{Haeni04, Biegalski05}, which would induce tensile strain
(a$_{film}$/a$_{substrate}$-1) of about 1.4 \% on the PTO film. The
cubic structure of PTO around \textit{T}$_C$= 490 $^{o}$C, has a
lattice parameter that is almost identical to the pseudocubic
lattice parameter of DSO \cite{Biegalski05} and at the deposition
temperature there is good lattice match between PTO and DSO, while
the mismatch between SRO and DSO is only $\approx 0.4$
\cite{Biegalski05}; this, in combination with the reduced thickness,
prevents the relaxation of either layer via misfit dislocations
\cite{Matthews74}. As it cools down below \textit{T}$_{C}$, however,
the drive towards the tetragonal structure introduces large in-plane
strains. These may be accommodated by dislocations
\cite{Matthews74}, by domain twinning \cite{Kwak, Speck94, Li01}
and/or by changes in the symmetry of the films\cite{Pertsev98,
Dieguez05}.

\begin{figure}[tbp]

\includegraphics[scale=0.4]{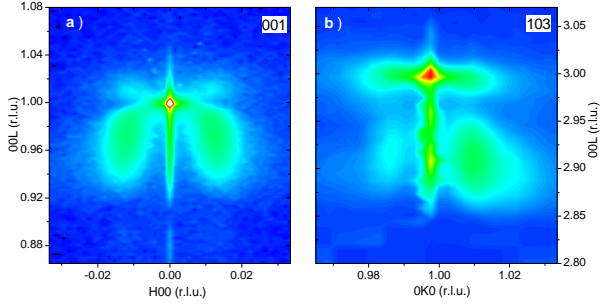}
\caption{(Color online)Logarithmic reciprocal maps in the H0L
scattering plane around the 001 (a) and 103 (b) reflections for a 5nm
PbTiO$_{3}$ film directly grown on a DyScO$_3$ substrate. (Intensities from low to high: blue-green-yellow-red-white)}
\end{figure}

The synchrotron x-ray measurements were carried out in the W1
beamline at HASYLAB (DESY-Hamburg), using a 1x1mm$^{2}$
monochromatic beam of 9.8 keV ($\lambda$ =1.26515{\AA }). Reciprocal space maps within the $\mathit{H}0\mathit{L%
}$ and 0$\mathit{KL}$ zones were obtained in standard
reflection geometry. Scans in the $\mathit{HK}0$ zone were achieved
using grazing incidence diffraction (GID), in which a
non-propagating evanescent wave is used to diffract within the plane
of the film. This is attained by making the x-ray beam reach the
surface close to the critical angle of total reflection (0.3$^{o}$
in our case). The reciprocal space maps presented here have been
calculated from the measured angular scans and plotted in reciprocal
lattice units (r.l.u.) of the substrate (1 r.l.u.$= 2\pi/3.950{\AA }$, in-plane, and $= 2\pi/3.942{\AA }$, out-of-plane).

\begin{figure}[tbp]

\includegraphics[scale=0.4]{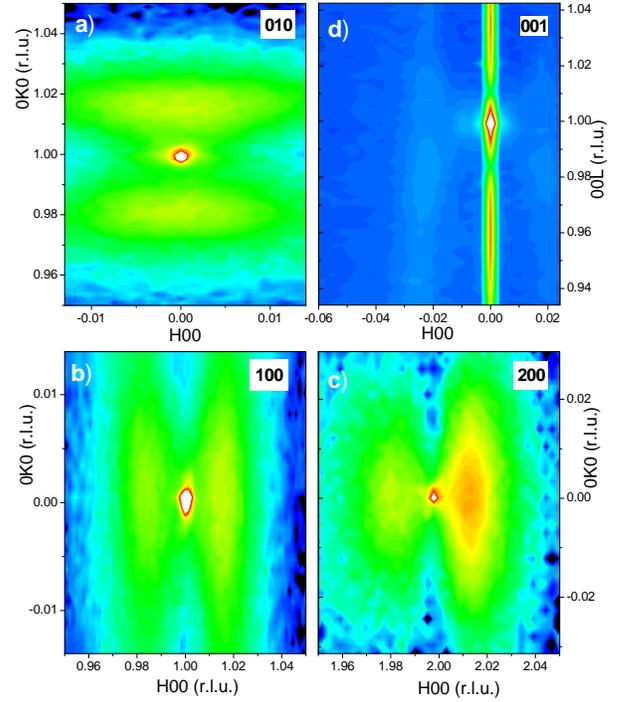}
\caption{(Color online) Logarithmic reciprocal space maps in the HK0
scattering zone around the a) 010, b)100 and c)200 reflections; and
in the H0L zone around the 001 reflection (d), for the film grown
with an SRO electrode. (Intensities from low to high: b-g-y-r-w) }
\end{figure}

The reciprocal space maps around the (001) and (103) reflections
(Figure 1) show that the in-plane reciprocal space vectors of the
film and the substrate (the sharp peaks at L=1 (a) and L=3(b)) are
identical (same H value in the figures), and that the films'
truncation rods have the same width as the substrate peak. These
observations are evidence of full in-plane coherence of the films.
However, Figure 1 also shows broad satellite peaks outside of the
truncation rod. In Figure 2 the GID reciprocal space maps within the
HK0 plane, corresponding to the (100), (010) and (200) in-plane
reflections of the film grown with a SRO buffer layer are shown,
together with the scan in the H0L plane around the specular 001
reflection of the same sample. For a given sample, the satellites
have the same reciprocal space separation in all reflections. This
shows that they are not Bragg peaks owing to different elastic
domains, or twins, but originate instead from an in-plane
modulation. The distance between the satellites and the Bragg peak
($\Delta \mathit{H}$) yields modulation periodicities of $\Lambda=
2\pi/\Delta \mathit{Q}= 0.395nm/\Delta\mathit{H}\simeq$ 20nm for the
films grown with an SRO electrode, and $\Lambda\simeq$ 30nm for the
films grown directly on DSO.

Figure 3 shows the HKO area scans around the 110 reflections of the
two films. While the satellites for the film grown on an SRO
electrode are oriented preferentially along the a, b pseudo-cubic
axes (the edges of the substrate), the intensity distribution for
the film grown directly on DSO is somewhat more isotropic (same
near-circular distribution with increased intensity at high H and K
values was found around the 100 and 010 reflections of the same
sample). We do not know the cause for this difference, although we
note that both types of domain configuration have been seen before
for PTO films grown on SrTiO$_3$
substrates\cite{Streiffer02,Fong04}.

\begin{figure}[tbp]

\includegraphics[scale=0.4]{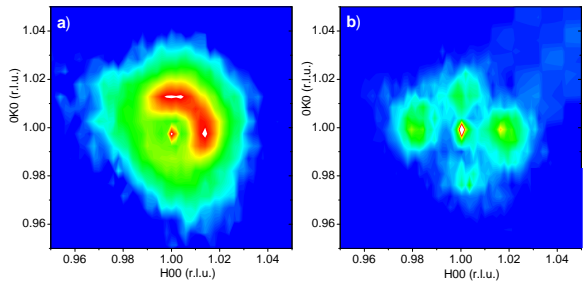}
\caption{(Color online)Logarithmic reciprocal space maps of the HK0
plane around the 110 Bragg peaks of the films grown a)directly on
DSO, and b)on DSO with an SRO bottom electrode. (Intensities from
low to high: b-g-y-r-w)}
\end{figure}

The pseudocubic lattice parameters of the DSO substrate were found to be a$%
\simeq$ b=3.950 {\AA }, and c=3.942 {\AA }, while the lattice
parameters of the PTO films were a$\simeq$b=3.950 {\AA }, owing to
the in-plane coherence, and c$\simeq$4.07{\AA }\cite{orthorhombic}.
As reported by He \textit{et al}. \cite{He05, He04}, the clamping to
the substrate can decouple the symmetry of the unit cell from that
of the polar shifts. Moreover, from a functional point of view, as
important as the crystal structure is the polar symmetry and its
domain configuration. In order to elucidate this, we turn to the
analysis of the satellite peaks.

In-plane satellites in x-ray diffraction patterns may have different
origins: grain boundaries, dislocations, disclinations, or other
periodic structural discontinuities, or else, in ferroelectric
materials, polar domains. We can disregard grain boundaries from AFM
analysis of the film surface \cite{Janssens06}. As for misfit
dislocations, the good lattice match during growth makes them
unlikely. In fact, the symmetry of the 100 and 010 satellites, along
[h00] and [0k0], respectively (Figure 2) is inconsistent with
scattering by dislocations, grain boundaries or any other periodic
defects, since they should render modulation along the same
directions in all reflections, which is not the case. Thus, the
satellites are not due to periodic defects, but to polar domains
\cite{Catalan06}. Indeed the in-plane diffraction patterns are
extremely similar to some of those observed around the 304
reflections for PTO films with vertical stripe domains in
refs.\cite{Streiffer02, Fong04}.

Stripes of alternate polarization form in a ferroelectric under
open-circuit electrical boundary conditions in order to minimize the
depolarizing field \cite{Streiffer02,Mitsui53} and/or the epitaxial
strain \cite{Speck94,Li01,Kornev04}. Since polarization is due to a shift of
the Pb and Ti cations with respect to the oxygen cage, the different
shift directions in the different domains produces contrast in the
structure factors and causes the intensity modulations. Accordingly,
there can only be structure-factor contrast (and therefore
satellites) around those reciprocal lattice vectors that have at
least one component parallel to the direction of the polar
displacement\cite{Fong04}. Thus, the distribution of the satellites
yields information about the domain geometry and size, and the Bragg
peaks around which the satellites appear tell us about the direction
of the polarization.

\begin{figure}[tbp]

\includegraphics[scale=0.45]{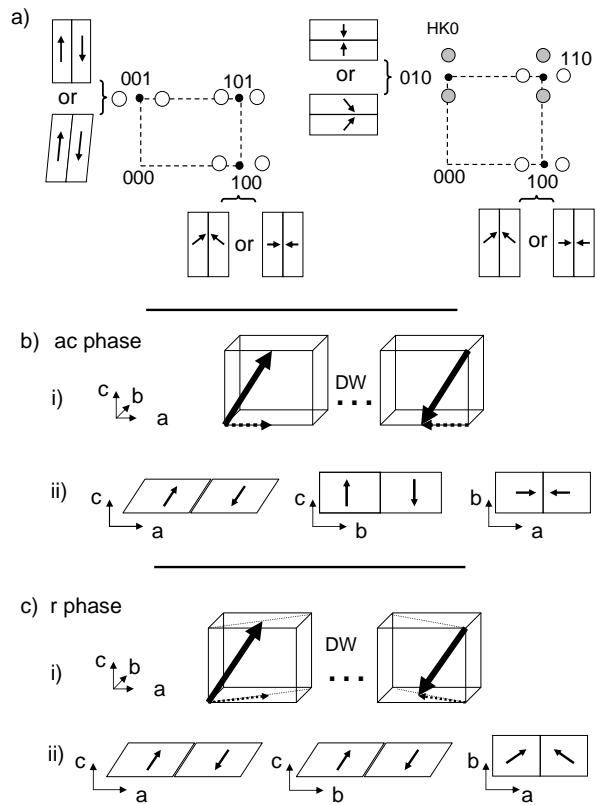}
\caption{a) Schematic representation of the observed reciprocal
space maps and their implication for the polar domain orientation;
Polar symmetry for b) $\textit{ac}$-phase, and c) the
$\textit{r}$-phase. In b)and c): i) sketches the polar shifts in two
unit cells at both sides of the domain wall and ii) represents the
projections of the polarization in the \{100\} pseudo-cubic planes
for the proposed phases and domain configurations.}
\end{figure}

Therefore, from Figs. 1-3, we learn: i) Polarization must have both
in-plane and out-of-plane components, since there are satellites
around (00l),(h0l),(h00),(0k0) and (hk0) reflections, all with the
same size, and ii) the in-plane components of the polarization must
be oriented head-to-head (and tail-to-tail) to produce modulation
parallel to the in-plane reciprocal lattice vectors. In Figure 4 we
have drawn a scheme of the domain orientations compatible with i) and ii). These are the \textit{ac} phase (P$%
_{x}$$\neq$ 0, P$_{y}$=0, P$_{z}$$\neq$ 0, monoclinic Pm) with in-plane
component of the polarization along the edges of the simple cubic cell (see
Fig. 4b), and the \textit{r}-phase (P$_{x}$=P$_{y}$$\neq$ P$_{z}$,
monoclinic space group Cm) with in-plane component of the polarization along
the face diagonals of the simple cubic cell (see Fig. 4c). To reproduce the observed maps in Fig. 4a,
we consider four-fold in-plane symmetry and the existence of domain walls at 0
and 90$^o$ in both phases.

It is worth noting that the domain walls responsible for the
in-plane satellites are charged, which is energetically costly.
Although this is a surprising result, there are precedents of
head-to-head polarized domains in other perovskites\cite{Little55,Cao}.
If the polarization initially nucleates randomly, when domains with opposite
polarization grow and encounter, whether or not they switch in order
to match will depend on the balance between the energy needed for
the switching and the amount of charge in the wall. If this charge
is neutralized by defects, or directly short-circuited by the
electrode, the domains will remain head-to-head. This picture is
consistent with the bigger size of the domains in the films grown
without SRO electrode. It is also worth noting that for ultra-thin
films subjected to partially-screened electric fields, the polar
orientation has been predicted to rotate between the bottom and the
top surface in such a way that in-plane head-to-head and
tail-to-tail polar states appear at the top and the bottom layers of
the film \cite{Kornev04}.

The out-of-plane lattice parameters are the same for the films grown
directly on DSO as for those grown with an SRO layer. Therefore, the
lower intensity of the satellites observed for the latter, specially
around the out-of-plane reflections (Figure 2-d) seems to indicate
not a smaller out-of-plane polarization but rather a lack of
modulation in the out-of-plane component of the polarization. This
agrees with previous observations pointing that a lower electrode
helps charge neutralization, thereby preventing the formation of
c-axis antiparallel stripes \cite{Thompson04}

In general our results agree with theoretical calculations that
predict the rotation of the polarization, from perpendicular to
parallel to the film, under tensile strain. However there is
discrepancy between the phases we proposed and those predicted for
PbTiO$_3$ films. An \textit{aa}-phase (P$_{x}$=P$_{y}$, P$_{z}$=0)
have been theoretically predicted for the present strain
state\cite{Pertsev98, Bungaro04, Dieguez05, Li01}, but it is
incompatible with the observed \textit{c}-axis polarization.
Combinations of \textit{aa}-phase and
\textit{c}-phase\cite{Dieguez05} or \textit{aa}-phase and
\textit{r}-phase\cite{Pertsev00} have also been proposed, but they
would imply two different out-out-plane lattice parameters, not
observed here. The \textit{r}-phase proposed by us has been
predicted by Landau models for PbTiO$_3$ under different
(smaller) strain states\cite{Pertsev98}, and by ab-initio
calculations for PTO under hydrostatic pressure\cite{Cohen05}. The
\textit{ac} state has never been predicted for PTO. This discrepancy
is in a sense not surprising, since most current theoretical
calculations do not include electrical boundary conditions and are
therefore not valid for very thin films. These boundary conditions
have already been shown to affect the phase diagram of BaTiO$_{3}$
\cite{Bellaiche05} and the domain structure of
PbZr$_{1-x}$Ti$_{x}$O$_{3}$ under compressive strain
\cite{Kornev04}.

In sum, fully coherent, very thin films of PbTiO$_{3}$ under tensile
strain display a modulation in reciprocal space indicating stripe
domains with the polarization rotated away from the tetragonal axis.
This new low-symmetry state may provide a route towards achieving
large piezoelectric coefficients in undoped PTO \cite{Cohen05}.

We thank L. Bellaiche, T. Hibma, I. Kornev, J.F. Scott, A. Vlooswijk and V.
Vonk for useful discussions and the Dutch organizations FOM and NWO
for financial support. G.C. acknowledges financial support from the
Marie-Curie Intra-European fellowship programme.


\begin{references}
\bibitem{Dawber} M. Dawber, K. M. Rabe, J. F. Scott, Rev. Mod. Phys. \textbf{77}, 1083
(2005).
\bibitem{Rossetti91} G. A. Rossetti, L. E. Cross, K. Kushida, Appl. Phys. Lett. \textbf{59},
2524 (1991).
\bibitem{Choi04}K. J. Choi, M. Biegalski, Y. L. Li, A. Sharan, J. Schubert, R.
Uecker, P. Reiche, Y. B. Chen, X. Q. Pan, V. Gopalan, L.-Q. Chen, D.
G. Schlom, C. B. Eom, Science \textbf{306}, 1005 (2004).
\bibitem{Haeni04} J. H. Haeni, P. Irvin, W. Chang, R. Uecker, P. Reiche, Y. L. Li,
S. Choudhury, W. Tian, M. E. Hawley, B. Craig, A. K. Tagantsev, X.
Q. Pan, S. K. Streiffer, L. -Q. Chen, S. W. Kirchoefer, J. Levy and
D. G. Schlom, Nature (London) \textbf{430}, 758 (2004).
\bibitem{He05}F. He, B. O. Wells, and S. M. Shapiro, Phys. Rev. Lett. \textbf{94},
176101 (2005).
\bibitem{Basceri97} C. Basceri, S. K. Streiffer, A. I. Kingon, R. Waser, J. Appl.
Phys. \textbf{82}, 2497 (1997).
\bibitem{Pertsev98} N. A. Pertsev, A. G. Zembilgotov, A. K. Tagantsev, Phys. Rev.
Lett. \textbf{80}, 1988 (1998).
\bibitem{Dieguez05} O. Diéguez, K. M. Rabe, and D. Vanderbilt, Phys. Rev. B \textbf{72},
144101 (2005).
\bibitem{Bungaro04}C. Bungaro and K. Rabe, Phys. Rev. B \textbf{69}, 184101 (2004)
\bibitem{Speck94} J. S. Speck and W. Pompe, J. Appl. Phys. \textbf{76}, 466 (1994); J. S.
Speck, A. C. Daykin, A. Seifert, A. E. Romanov and W. Pompe, J.
Appl. Phys. J. Appl. Phys. \textbf{78}, 1696 (1995).
\bibitem{Li01} Y.L. Li, L.Q. Chen, Appl. Phys. Lett. \textbf{88}, 072905 (2006).
\bibitem{Kornev04} I. Kornev, H. Fu, and L. Bellaiche, Phys. Rev.
Lett. \textbf{93}, 196104 (2004).
\bibitem{Kwak} B. S. Kwak, A. Erbil, J. D.
Budai, M. F. Chisholm et al. Phys. Rev. Lett. \textbf{68}, 3733 (1992)
\bibitem{Stemmer} S. Stemmer, S.K. Streiffer, F. Ernst, M. Ruhle, W-Y. Hsu and R. Raj, Solid State Ionics \textbf{75}, 43 (1995)
\bibitem{Biegalski05} M.D. Biegalski, J.H. Haeni, S. Trolier-McKinstry, D.G. Schlom,
C.D. Brandle, and A.J. Ven Graitis, J. Mat. Res. \textbf{20}, 952
(2005).
\bibitem{Kobayashi83} Although a small -and unconfirmed- orthorhombic distortion
was reported for bulk PTO below -90$^{o}$C, the polar direction
remained c-axis at all temperatures (J. Kobayashi \emph{et al.} , Phys. Rev. B \textbf{28}, 3866 (1983)).
\bibitem{Noheda00} R. Guo, L. E. Cross, S-E. Park, B. Noheda, D. E. Cox, G. Shirane, Phys. Rev. Lett. \textbf{84}, 5423 (2000).
\bibitem{Bellaiche00} L. Bellaiche, A. Garcia, D. Vanderbilt, Phys. Rev. Lett. \textbf{84}, 5427 (2000).
\bibitem{Junquera03} J. Junquera, P. Ghosez, Nature 422, 506 (2003).
\bibitem{Streiffer02} S. K. Streiffer, J. A. Eastman, D. D. Fong, C. Thompson, A.
Munkholm, M.V. R. Murty, O. Auciello, G. R. Bai, and G. B.
Stephenson, Phys. Rev. Lett. \textbf{89}, 67601 (2002).
\bibitem{Fong04}14. D. Fong, G. Brian Stephenson, S. K. Streiffer, J. A. Eastman, O.
Auciello, P. H. Fuoss, C. Thompson, Science \textbf{304}, 1650 (2004).
\bibitem{Triscone05}L. Despont, C. Lichtensteiger, C. Koitzsch, F. Clerc, M. G. Garnier, F. J.
Garcia de Abajo, E. Bousquet, Ph. Ghosez, J.-M. Triscone, and
P.Aebi, arXiv:cond-mat/0511084 (2005).
\bibitem{Janssens06}A. Janssens \textit{et al.}, in preparation.
\bibitem{Rijnders} A.J.H.M. Rijnders, G. Koster, D.H.A. Blank, H. Rogalla,
Appl. Phys. Lett. \textbf{70}, 1888 (1997)
\bibitem{Shirane50} G. Shirane, S. Hoshino, K. Suzuki, Phys. Rev. \textbf{80}, 1105 (1950).
\bibitem{Matthews74}J.W. Matthews, A.E. Blakeslee, J. Cryst. Growth \textbf{27}, 118 (1974).
\bibitem{orthorhombic}In some substrates we measured a=b=3.950{\AA }
and in others a= 3.950{\AA }, b=3.960{\AA } (see Fig. 3 a) and b).
On the latter ones the PTO unit cell cannot not have the tetragonal
symmetry of the bulk phase and is instead orthorhombic. The values
measured by us differ slightly from those reported in the
literature.
\bibitem{He04}F. He, B. O. Wells, Z. G. Ban, S. P. Alpay, S. Grenier, S. M. Shapiro, W. Si, A. Clark, and X. X. Xi, Phys. Rev. B \textbf{70}, 235405 (2004).
\bibitem{Catalan06} Further evidence for a polar origin of the satellites was provided
by their vanishing upon heating above \textit{T}$_{C}$ in thicker (30nm) films(G. Catalan \textit{et al.}, in preparation).
\bibitem{Mitsui53} T. Mitsui, J. Furuichi, Phys. Rev. \textbf{90}, 193 (1953).
\bibitem{Little55}E.A. Little, Phys. Rev. \textbf{98}, 978 (1955).
\bibitem{Cao} Y. Jin and W. Cao, Appl. Phys. Lett. \textbf{87}, 7438 (2000); J. Han and W. Cao, Appl. Phys.
Lett. \textbf{83}, 2040 (2003).
\bibitem{Thompson04} C. Thompson, communication, Intal. Meeting on Fundamental Physics of Ferroelectrics, Williamsburg 2004.
\bibitem{Pertsev00} N. A. Pertsev, V. G. Koukhar, Phys. Rev. Lett. \textbf{84}, 3722 (2000).\bibitem{Cohen05} Z. Wu and R.E. Cohen, Phys. Rev. Lett. \textbf{95}, 037601 (2005)
\bibitem{Bellaiche05} Bo-Kuai Lai, Igor A. Kornev, L. Bellaiche, and G. J.
Salamo, Appl. Phys. Lett. \textbf{86}, 132904 (2005).


\end{references}
\end{document}